\documentclass[superscriptaddress,pre,aps,twocolumn,showpacs,floatfix]{revtex4}

\usepackage[dvips]{graphicx}
\usepackage{amsmath}
\usepackage{amssymb}
\usepackage[nice]{nicefrac}
\usepackage{color}
\usepackage{amsmath}
\usepackage{grffile}
\usepackage{times}

\newcommand{\pst}{p_{\mathrm{st}}}

\newcommand{\sgn}{\mbox{sign}}

\renewcommand{\=}{\stackrel{d}{=}}

\begin{document}
\title{Stationary states in 2D systems driven by bi-variate L\'evy noises}

\author{Krzysztof Szczepaniec}
\email{kszczepaniec@th.if.uj.edu.pl}
\affiliation{Marian Smoluchowski Institute of Physics, and Mark Kac Center for Complex Systems Research, Jagiellonian University, ul. Reymonta 4, 30--059 Krak\'ow, Poland }

\author{Bart{\l}omiej Dybiec}
\email{bartek@th.if.uj.edu.pl}
\affiliation{Marian Smoluchowski Institute of Physics, and Mark Kac Center for Complex Systems Research, Jagiellonian University, ul. Reymonta 4, 30--059 Krak\'ow, Poland }

\date{\today}

\begin{abstract}
Systems driven by $\alpha$-stable noises could be very different from their Gaussian counterparts. Stationary states in single-well potentials can be multimodal.  
Moreover, a potential well needs to be steep enough in order to produce stationary states.
Here, it is demonstrated that 2D systems driven by bi-variate $\alpha$-stable noises are even more surprising than their 1D analogs.
In 2D systems, intriguing properties of stationary states originate not only due to heavy tails of noise pulses, which are distributed according to $\alpha$-stable densities, but also because of properties of spectral measures. Consequently, 2D systems are described by a whole family of Langevin and fractional diffusion equations. Solutions of these equations bear some common properties but also can be very different.
It is demonstrated that also for 2D systems potential wells need to be steep enough in order to produce bounded states.
Moreover, stationary states can have local minima at the origin. The shape of stationary states reflects symmetries of the underlying noise, i.e. its spectral measure.
Finally,  marginal densities in power-law potentials also have power-law asymptotics.

\end{abstract}

\maketitle

\section{Introduction\label{sec:introduction}}

The description of complex interactions with surrounding can be significantly simplified by a use of the noise, which provides effective approximation to not fully known processes.
Description of many systems can be provided by the Langevin equation, which in the overdamped limit typically is of the form
\begin{equation}
\dot{x}(t)=f(x)+\zeta(t),
\label{eq:langevin0}
\end{equation}
where $f(x)$ is the deterministic force, while $\zeta(t)$ represents complex interactions of the test ``particle'' with its environment.
Usually it is assumed that the noise is white and Gaussian.
White type of the noise underlines that stochastic pulses acting on a test particle are independent.
The Gaussianity is the consequence of the central limit theorem, which states that a distribution of a sum of many independent and bounded summands converges to the normal (Gaussian) distribution.
If the assumption about bounded type of interactions is relaxed, the noise can be still  of the white type but it becomes of more
general L\'evy type by virtue of the generalized central limit theorem \cite{gnedenko1968,samorodnitsky1994}.
The Langevin equation~(\ref{eq:langevin0}) underlines many concepts and problems of statistical physics and theory of stochastic process like resonant activation \cite{doering1992} or stochastic resonance \cite{gammaitoni1998,anishchenko1999} to name a few.

Analogously like the white Gaussian noise plays special role in the theory of stochastic systems also the L\'evy type noise becomes more important as it is capable to describe out-of-equilibrium phenomena.
The special role played by L\'evy stable distributions is due to their inherent properties: stability (invariance under convolution),
power law asymptotics and the generalized central limit theorem. Consequently, stable distributions provide general, well developed framework for description of many out-of-equilibrium phenomena revealing large bursts, outliers and asymmetry \cite{Weeks1998}. This more general framework incorporates Gaussian realms as a special case.
The presence of fluctuations distributed according to L\'evy laws have been observed in various situations in
physics, chemistry or biology \cite{shlesinger1995,nielsen2001},
paleoclimatology \cite{ditlevsen1999b} or economics
\cite{mantegna2000}. The heavy-tailed
fluctuations appear in context of different models
\cite{solomon1993,chechkin2002b,boldyrev2003}, and are
analyzed in an increasing number of studies
\cite{jespersen1999,ditlevsen1999,dybiec2004,dybiec2006b,chechkin2002,chechkin2003,dubkov2008,garbaczewski2009,srokowski2009b,rypdal2010,barthelemy2008,metzler2007,klages2008,dybiec2009e,zeng2010}.
Despite some controversies regarding observability of L\'evy flights \cite{Edwards2007,gonzalez2008} and theoretical issues, e.g. unbounded variance, L\'evy flights are considered as a paradigm of  optimal search strategies among randomly distributed target sites \cite{Viswanathan1996}.

The present work addresses properties of 2D L\'evy flights in external
potentials. Theoretical descriptions of such systems are based on the
Langevin equation and/or fractional Smoluchowski-Fokker-Planck equation.
The research performed here extends earlier studies of 1D systems
\cite{jespersen1999,chechkin2002,chechkin2003,chechkin2004,dubkov2005b,dubkov2007,samorodnitsky2003,samorodnitsky2007,dybiec2007d,dybiec2010d} where
analysis of symmetric and asymmetric L\'evy flights in harmonic, superharmonic and subharmonic potentials
have been presented.
The discussion conducted there covered the problem of existence and properties of stationary states in 1D systems.
The current studies extend the  description of multi-variate systems perturbed by L\'evy flights \cite{chechkin2000b,chechkin2002b} and comment on various types of fractional Smoluchowski-Fokker-Planck equations \cite{yanovsky2000,schertzer2001} associated with 2D L\'evy flights, depending on the type of the spectral measure \cite{samorodnitsky1994}.
Here, characteristics of 2D L\'evy flights in confining potentials are studied for various types of bi-variate $\alpha$-stable motions.
Properties of stationary states in 2D potentials are inspected by the use of analytical arguments and Monte Carlo simulations.

The studied model is presented in Section~\ref{sec:model}.
Section~\ref{sec:results} discusses obtained, mainly numerically, results.
The paper is closed with concluding remarks (Section~\ref{sec:summary}).

\section{Model \label{sec:model}}

The main scope of the current article is to study 2D systems driven by bi-variate $\alpha$-stable noise.
Properties of bi-variate $\alpha$-stable noises are very different from their 1D analogs.
Thus, the basic information about bi-variate $\alpha$-stable random variables and associated $\alpha$-stable motions is provided, see Sec.~\ref{sec:bivariate}.
Next, the model of a 2D generalized random walk in spherically symmetric 2D potentials subject to bi-variate $\alpha$-stable noises is presented, see Sec.~\ref{sec:motivation}.

\subsection{Bi-variate $\alpha$-stable motion\label{sec:bivariate}}

Increments of the 1D Wiener (Brownian motion) process are distributed according to the Gaussian distribution. There are two natural generalizations of the Wiener process. The first option is to study multi dimensional Wiener process. The next option is to relax an assumption about Gaussianity of increments. Instead of normal density it is possible to assume that increments are distributed according to the more general $\alpha$-stable densities \cite{samorodnitsky1994,janicki1994}, which include the normal distribution as a special case. The later extension defines an $\alpha$-stable L\'evy type motion. Here, we are interested in properties of the 2D system driven by bi-variate L\'evy noise which is a formal time derivative of the bi-variate L\'evy motion. Such an extension unifies both aforementioned options --- simultaneously system departs from the Gaussianity and its dimensionality increases.

For the clarity of presentation we give a pedagogical introduction to the required theory. Following \cite{samorodnitsky1994} we present basic definitions and properties of $\alpha$-stable densities. Next, we show how multi-variate L\'evy motion can be defined. Finally, using presented methods we 
will 
study properties of  stationary states in 2D overdamped systems.

Multidimensional stable random variables are defined in the analogous way like 1D stable random variables. 
A random vector $\boldsymbol{X}=(X_1,\dots,X_d)$ is said to be a stable random vector in $\mathbb{R}^d$  if for any positive numbers $A$ and $B$, there is a positive number $C$ and a vector  $\boldsymbol{D}$ such that
\begin{equation}
 A\boldsymbol{X}^{(1)}+B\boldsymbol{X}^{(2)}  \= C\boldsymbol{X} + \boldsymbol{D},
 \label{eq:definition}
\end{equation}
where $\boldsymbol{X}^{(1)}$ and $\boldsymbol{X}^{(2)}$ are independent copies of $\boldsymbol{X}$, $\=$ denotes equality in distributions. The vector $\boldsymbol{X}$ is called strictly stable if $\boldsymbol{D}=\boldsymbol{0}$. The vector $\boldsymbol{X}$ is symmetric stable if it is stable and satisfies an additional relation
\begin{equation}
\mathrm{Prob}\{\boldsymbol{X} \in A \} = \mathrm{Prob}\{-\boldsymbol{X} \in A \} 
\end{equation}
for any Borel set $A$ of $\mathbb{R}^d$.
A random vector is $\alpha$-stable if Eq.~(\ref{eq:definition}) holds with $C=(A^\alpha+B^\alpha)^{1/\alpha}$ where $0 < \alpha \leqslant 2$.

The characteristic function of $\alpha$-stable densities can be determined by defining Eq.~(\ref{eq:definition}). In 1D the characteristic function is \cite{samorodnitsky1994,janicki1994}
\begin{equation}
 \phi(k)=  \mathbb{E} \left[ e^{ikX} \right] = 
 \left\{
 \begin{array}{l}
 \exp\left[  -\sigma^\alpha |k|^\alpha \left( 1 - i\beta\sgn k \tan\frac{\pi\alpha}{2}  \right) +i\mu k \right] \\
 \;\;\;\;\;\;\;\;\;\;\mbox{if}\;\;\alpha\neq 1, \\
  \exp\left[  -\sigma |k| \left( 1 + i\beta\frac{2}{\pi}\sgn k \ln |k|  \right) + i\mu k \right]  \\
 \;\;\;\;\;\;\;\;\;\;\mbox{if}\;\;\alpha= 1,
 \end{array}
 \right.
 \label{eq:characteristic1d}
\end{equation}
where $\alpha \in (0,2]$ is the stability index, $\beta \in [-1,1]$ is the asymmetry (skewness) parameter, $\sigma > 0$ is the scale parameter and finally $\mu \in \mathbb{R}$ is the location parameter. The closed formulas for $\alpha$-stable densities are known only in a limited number of cases: $\alpha=2$ --- normal distribution, $\alpha=1$ with $\beta=0$ --- Cauchy distribution and $\alpha=\nicefrac{1}{2}$ with $\beta=1$ --- L\'evy-Smirnoff distribution.
In general, symmetric $\alpha$-stable densities with $\alpha<2$ have the power-law asymptotics of $|x|^{-(\alpha+1)}$ type.
If $X$ is the $\alpha$-stable random variable its characteristic function is given by Eq.~(\ref{eq:characteristic1d}). The fact that $X$ is distributed according to the $\alpha$-stable density is  schematically denoted by $X \sim S_\alpha(\sigma,\beta,\mu)$.

The characteristic function $\phi(\boldsymbol{k}) = \mathbb{E} \left[ e^{i \langle \boldsymbol{k},\boldsymbol{X}\rangle} \right]$ of the $\alpha$-stable vector $\boldsymbol{X}=(X_1,\dots,X_d)$ in $\mathbb{R}^d$ is given by \cite{samorodnitsky1994}
\begin{widetext}

\begin{equation}
 \phi(\boldsymbol{k}) =
 \left\{
 \begin{array}{lcl}
   \exp\left\{ -\int_{S_d}  |\langle \boldsymbol{k},\boldsymbol{s} \rangle|^\alpha  \left[ 1 -i\sgn(\langle \boldsymbol{k},\boldsymbol{s} \rangle)\tan\frac{\pi\alpha}{2} \right]\Gamma(d\boldsymbol{s}) +i \langle \boldsymbol{k},\boldsymbol{\mu}^0 \rangle  \right\} & \mbox{for} & \alpha\neq 1,\\
      \exp\left\{ -\int_{S_d}  |\langle \boldsymbol{k},\boldsymbol{s} \rangle|^\alpha  \left[ 1 +i\frac{2}{\pi}\sgn(\langle \boldsymbol{k},\boldsymbol{s} \rangle)\ln(\langle \boldsymbol{k},\boldsymbol{s} \rangle) \right]\Gamma(d\boldsymbol{s}) +i \langle \boldsymbol{k},\boldsymbol{\mu}^0 \rangle  \right\} & \mbox{for} & \alpha = 1,
  \end{array}
 \right.
 \label{eq:characteristicdd}
\end{equation}

\end{widetext}
where $\langle \boldsymbol{k} , \boldsymbol{s} \rangle$ represents the scalar product, $\Gamma(\cdot)$ stands for the (finite) spectral measure on the unit sphere  $S_d$ of $\mathbb{R}^d$ and $\boldsymbol{\mu}^0$ is a vector in $\mathbb{R}^d$, see \cite{samorodnitsky1994}. For any $d$-dimensional $\alpha$-stable vector any linear combination of its components in an $\alpha$-stable random variable.

The spectral measure $\Gamma(\cdot)$ contains information about symmetry (skewness) and scale of the probability density. More precisely, spectral measure $\Gamma(\cdot)$ replaces parameters $\beta$ and $\sigma$ which characterize 1D $\alpha$-stable densities, see Eq.~(\ref{eq:characteristic1d}). Multi-variate $\alpha$-stable density is said to be symmetric if the spectral measure is symmetric.

Example 2.3.3 from \cite{samorodnitsky1994} demonstrates how $\sigma$ and $\beta$ can be calculated in 1D  from the spectral measure. In 1D $S_1$ consists of two points $\{-1\}$ and $\{1\}$ and the spectral measure is concentrated on them. The characteristic function~(\ref{eq:characteristicdd}) reduces to the characteristic function of the $\alpha$-stable density $S_\alpha(\sigma,\beta,\mu)$ with
\begin{equation}
 \sigma=(\Gamma(1)+\Gamma(-1))^{1/\alpha},
\end{equation}
\begin{equation}
 \beta=\frac{\Gamma(1)-\Gamma(-1)}{\Gamma(1)+\Gamma(-1)}
\end{equation}
and $\mu=\mu^0$, where $\Gamma(\pm 1)=\Gamma(\{\pm 1\})$.

In general, components of multi-variate $\alpha$-stable variables are dependent. An $\alpha$-stable random vector $\boldsymbol{X}=(X_1,\dots,X_d)$ has independent components if and only if its spectral measure is discrete and concentrated on the intersection of the axes with the unit sphere $S_d$, see \cite[Example 2.3.5]{samorodnitsky1994}. Consequently, in 2D the spectral measure is concentrated on the points $(1,0)$, $(0,1)$, $(-1,0)$ and $(0,-1)$, see also \cite{chechkin2000b,chechkin2002b}.
For clarity of presentation, we  will refer to the bi-variate $\alpha$-stable noise with such a spectral measure as discrete or symmetric discrete.

The difference between various choices of spectral measures can be easily seen in the Cauchy case, i.e. for $\alpha=1$. If the spectral measure is symmetric and concentrated on intersections of the axes with the unit sphere $S_2$ the bi-variate Cauchy distribution is
\begin{equation}
  p( x,  y  ) = \frac{1}{\pi}\frac{\sigma}{( x^2+\sigma^2)}\times \frac{1}{\pi}\frac{\sigma}{( y^2+\sigma^2)}.
  \label{eq:discretecauchy}
 \end{equation}
If the spectral measure is continuous and uniform, the 2D Cauchy distribution is
\begin{equation}
  p( x,  y  ) = \frac{1}{2\pi}\frac{\sigma}{( x^2+ y^2+\sigma^2)^{3/2}} .
  \label{eq:spehricalcauchy}
 \end{equation}

General $d$-dimensional $\alpha$-stable random variables can be generated according to \cite{nolan1998b}
\begin{equation}
 Z=
 \left\{
 \begin{array}{lcl}
  \sum_{j=1}^n \gamma_j^{1/\alpha} Z_j \boldsymbol{s}_j & \mbox{for} & \alpha \neq 1 \\
  \sum_{j=1}^n \gamma_j \left[  Z_j + \frac{2}{\pi} \ln \gamma_j \right] \boldsymbol{s}_j & \mbox{for} & \alpha = 1 \\
 \end{array}
\right.,
\label{eq:algorithm}
\end{equation}
where $Z_1,\dots,Z_n$ are independent identically distributed totally skewed, standardized 1D $\alpha$-stable random variables, i.e. $Z_i \sim S_\alpha(1,1,0)$, $\boldsymbol{s}_j \in S_d$.
Pseudo random numbers distributed according to the $\alpha$-stable densities can be generated according to well known formulas \cite{chambers1976,weron1996}.
The special cases of symmetric $\alpha$-stable densities with uniform or symmetric continuous spectral measures can also be generated using sub-Gaussian random vectors \cite{samorodnitsky1994}.


The properties of the $\alpha$-stable motion are determined by properties of the multi-variate $\alpha$-stable densities.  The bi-variate L\'evy motion can be generated using~(\ref{eq:algorithm}) but there are also other approximated methods which are able to reconstruct main properties of the process \cite{teuerle2009} and their efficiency rely on the generalized central limit theorem.
For a very related discussion see \cite{teuerle2012}.

\subsection{Motivation \& Equations\label{sec:motivation}} 

A motion of an overdamped particle subject to the $\alpha$-stable L\'evy type noise is described by the following Langevin equation
\begin{equation}
 \frac{dx}{dt} =  -V'(x) + \sigma\zeta_{\alpha,0}(t),
 \label{eq:langevin}
\end{equation}
which can be rewritten as
\begin{equation}
 dx = -V'(x)dt + \sigma dL_{\alpha,0}(t),
 \label{eq:langevin2}
\end{equation}
where $L_{\alpha,0}(t)$ is a symmetric $\alpha$-stable motion \cite{janicki1994}.
$\zeta_{\alpha,0}(t)$ represents a white $\alpha$-stable noise which is a formal time derivative of the symmetric $\alpha$-stable motion $L_{\alpha,0}(t)$.
Eq.~(\ref{eq:langevin}) is associated with the following fractional Smoluchowski-Fokker-Planck equation
 \begin{eqnarray}
 \label{eq:ffpe}
  \frac{\partial p(x,t)}{\partial t} & =  & \frac{\partial }{\partial x}\left[ V'(x)p(x,t) \right] + \sigma^\alpha \frac{\partial^{\alpha} p(x,t)}{\partial |x|^\alpha} \\
  & = & \frac{\partial }{\partial x}\left[ V'(x)p(x,t) \right] - \sigma^\alpha (-\Delta)^{\alpha/2} p(x,t), \nonumber
\end{eqnarray}
where $\frac{\partial^\alpha}{\partial |x|^\alpha}=-(-\Delta)^{\alpha/2}$ is the fractional Riesz-Weil derivative (laplacian) defined via its Fourier transform \cite{samko1993}
 \begin{equation}
  \mathcal{F}\left[ \frac{\partial^\alpha p(x,t)}{\partial |x|^\alpha}  \right] = \mathcal{F}\left[ -(-\Delta)^{\alpha/2} p(x,t)  \right] = -|k|^\alpha \mathcal{F}\left[ p(x,t) \right].
  \label{eq:weil}
 \end{equation}
 The stationary states for Eq.~(\ref{eq:ffpe}), can be calculated by setting the left hand side of Eq.~(\ref{eq:ffpe}) to zero. The stationary solutions exist for potential wells which are steep enough \cite{dybiec2010d} and are not of the Boltzmann-Gibbs type \cite{eliazar2003}. For $\alpha=2$ and any $V(x)$ such that $\lim_{|x|\to \infty} V(x)=+\infty$, stationary states exist and are of the Boltzmann-Gibbs type, i.e. $\pst(x) \propto \exp\left[  - V(x)/\sigma^2 \right]$. For $\alpha<2$, the exponent $c$ characterizing a potential well $V(x)=|x|^c$ needs to be larger than $2-\alpha$, i.e. $c>2-\alpha$, see \cite{dybiec2010d}. Otherwise, the potential well is not steep enough in order to produce stationary states. The very interesting situation takes place for $c=2$ when the stationary density is of the same type (except the scale parameter) as the $\alpha$-stable distribution associated with the underlying noise, see Eq.~(\ref{eq:langevin}) and \cite{chechkin2003,chechkin2004,chechkin2006}.
 This very special situation is the consequence of the linearity of Eq.~(\ref{eq:langevin}).

Analytical formulas for stationary states for systems driven by $\alpha$-stable noises with $\alpha<2$ are known only in very limited number of cases.
For the quartic 1D Cauchy oscillator, i.e.  $V(x)=\frac{1}{4}x^4$ with $\alpha=1$, the stationary state of the fractional diffusion equation~(\ref{eq:ffpe}) is given by \cite{chechkin2003,chechkin2004,chechkin2006}
\begin{equation}
p_{\mathrm{st}}(x) = \frac{\sigma}{\pi(\sigma^{4/3}-\sigma^{2/3}x^2+x^4)}.
\label{eq:1d-quartic}
\end{equation}
This distribution $\pst(x)$ is symmetric, characterized by a finite variance, power-law asymptotics and it is bi-modal with the minimum at the origin and two global maxima at $x=\pm \sigma^{1/3}/\sqrt{2}$.

By analogy to 1D system, the bi-variate system is described by the following Langevin equation driven by the bi-variate $\alpha$-stable L\'evy type noise
\begin{equation}
 \frac{d\boldsymbol{r}}{dt} = -\nabla V(\boldsymbol{r})  + \sigma\boldsymbol{\zeta}_{\alpha}(t).
 \label{eq:langevin2d}
\end{equation}
Eq.~(\ref{eq:langevin2d}) can be rewritten as
\begin{equation}
 d\boldsymbol{r} = -\nabla V(\boldsymbol{r}) dt + \sigma d\boldsymbol{L}_{\alpha}(t),
 \label{eq:langevin2d2}
\end{equation}
where $\boldsymbol{L}_{\alpha}(t)$ is a bi-variate $\alpha$-stable motion and $V(\boldsymbol{r})$ is an external potential.
By analogy to 1D cases multi-variate $\alpha$-stable noise $\boldsymbol{\zeta}_{\alpha}(t)$ is a formal time derivative of the multi-variate $\alpha$-stable motion $\boldsymbol{L}_{\alpha}(t)$.
Especially interesting potentials are harmonic $V(x,y)=\frac{1}{2}r^2=\frac{1}{2}(x^2+y^2)$ and quartic $V(x,y)=\frac{1}{4}r^4=\frac{1}{4}(x^2+y^2)^2$ which will be used in further investigations.

\begin{figure}[!ht]
 \includegraphics[angle=0,width=0.98\columnwidth]{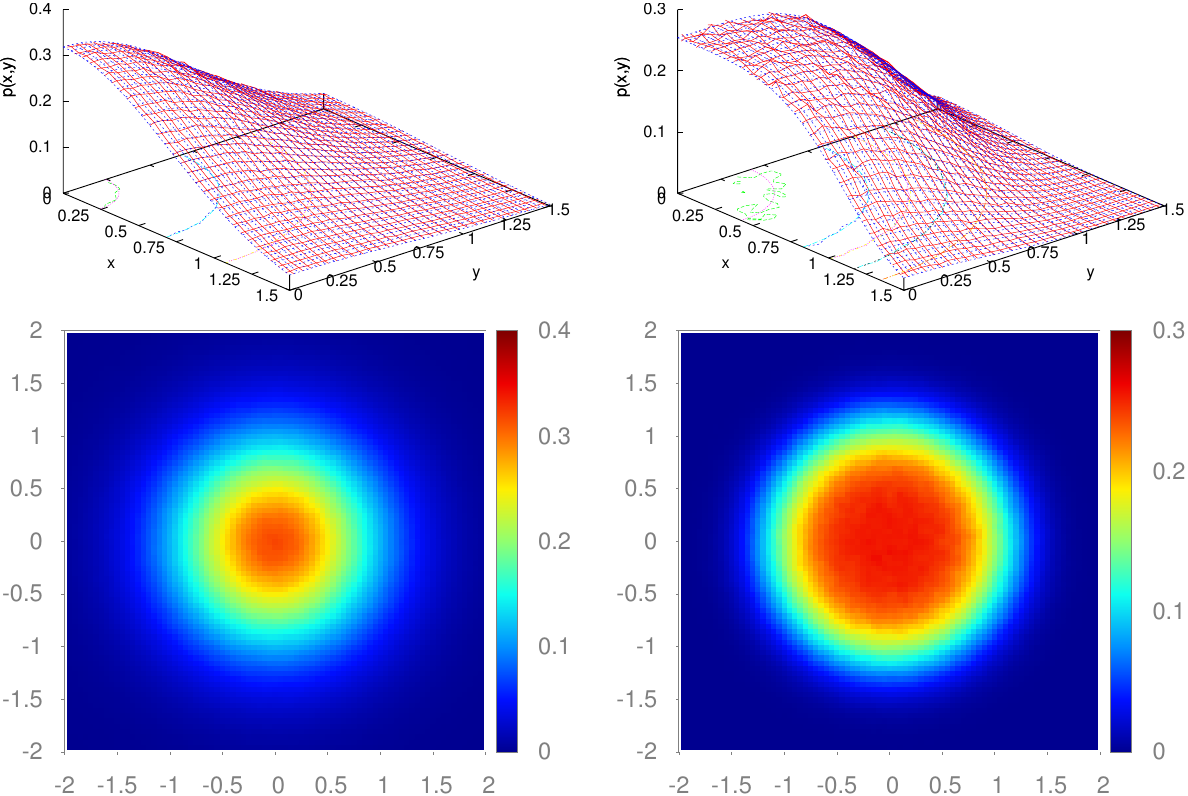}
 \caption{(Color online) 
 Stationary states for the harmonic potential $V(x,y)=\frac{1}{2}(x^2+y^2)$ (left panel) and the quartic potential $V(x,y)=\frac{1}{4}(x^2+y^2)^2$ (right panel) subject to bi-variate, uniform Gaussian white noise. Top row presents stationary densities as 3D surfaces while bottom row as heat maps.
 Theoretical stationary densities are presented as blue surfaces in the top row, while contours present theoretical and estimates isolines.
 The stationary densities have been estimated from the sample of $N=10^8$ elements with the integration time step $\Delta t=10^{-3}$.
}
\label{fig:parabolic-quartic-gauss}
\end{figure}

 The Langevin equation~(\ref{eq:langevin2d}) can be associated with the Smoluchowski-Fokker-Planck equation which has the general form
 \begin{equation}
  \frac{\partial p(\boldsymbol{r},t)}{\partial t} =
\nabla \cdot \left[ \nabla V(\boldsymbol{r}) p(\boldsymbol{r},t) \right]
  + \sigma^\alpha\boldsymbol{\Xi} p(\boldsymbol{r},t),
    \label{eq:ffpe2d}
 \end{equation}
 where $\boldsymbol{\Xi}$ is the fractional operator  due to bi-variate $\alpha$-stable noise $\boldsymbol{\zeta}$, see Eq.~(\ref{eq:langevin2d}), while $\nabla \cdot \left[ \nabla V(\boldsymbol{r}) p(\boldsymbol{r},t) \right]$ originates due to the deterministic force $\boldsymbol{F}(\boldsymbol{r})=-\nabla V(\boldsymbol{r})$ acting on a test particle.
In Eq.~(\ref{eq:ffpe2d}) the drift term has the standard form but the diffusive term depends on the noise type.
 For the bi-variate $\alpha$-stable noise with the uniform spectral measure $\boldsymbol{\Xi}=-(-\Delta)^{\alpha/2}$, i.e. it is the fractional laplacian defined via the Fourier transform, see \cite{samko1993}
 \begin{equation}
  \mathcal{F}\left[ -(-\Delta)^{\alpha/2} p(\boldsymbol{r},t)  \right] = -|\boldsymbol{k}|^\alpha \mathcal{F}\left[ p(\boldsymbol{r},t) \right].
  \label{eq:weil2d}
 \end{equation}
 For the bi-variate $\alpha$-stable noise with the discrete symmetric  spectral measure (located at intersections of axes with the unit sphere $S_2$) the fractional operator $\boldsymbol{\Xi}$ has the form
 $\boldsymbol{\Xi}=\frac{\partial^\alpha }{\partial |x|^\alpha} + \frac{\partial^\alpha }{\partial |y|^\alpha}$, where fractional derivatives are defined as in the 1D case. Such a case corresponds to the situation when jumps induced by the stochastic force along both axes are independent.
 This demonstrates that the Langevin equation is associated with the whole family of fractional Smoluchowski-Fokker-Planck equations depending on the noise type. The choice of the spectral measure $\Gamma(\cdot)$ determines the fractional operator $\boldsymbol{\Xi}$.

\begin{figure}[!ht]
\includegraphics[angle=0,width=0.98\columnwidth]{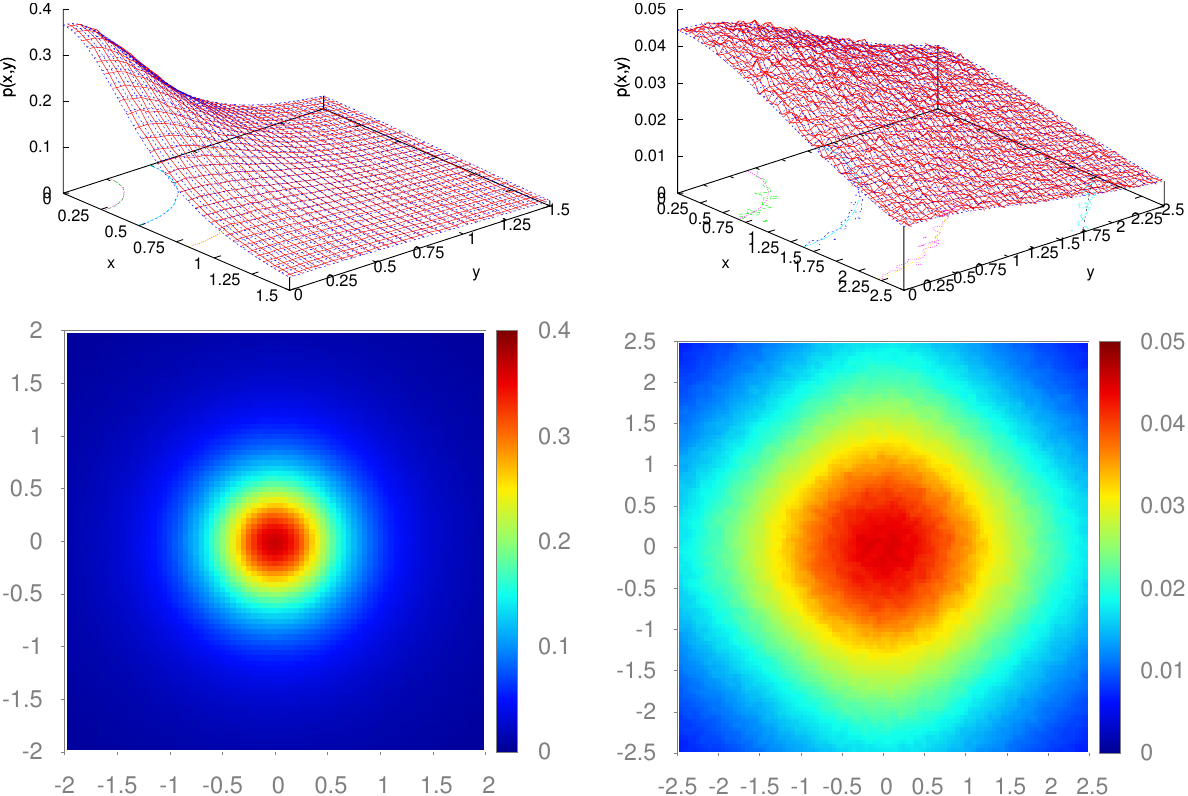}
 \caption{(Color online) Stationary states for the harmonic potential $V(x,y)=\frac{1}{2}(x^2+y^2)$  subject to the bi-variate Cauchy noises:
 spherical (uniform) Cauchy noise (left panel) and discrete Cauchy noise (right panel). Top row presents stationary probability densities as 3D surfaces while bottom row as heat maps.
 Theoretical stationary densities are presented as blue surfaces in the top row, while contours present theoretical and estimates isolines.
 The stationary densities have been estimated from the sample of $N=10^8$ elements with the integration time step $\Delta t=10^{-3}$.
}
\label{fig:harmonic--cauchy}
\end{figure}

The main scope of current research is to check if stationary states for harmonic and quartic potentials subject to bi-variate $\alpha$-stable noises exist and what are their shapes depending on a noise type.
In particular, we focus on two cases:
 (\textit{a}) stable bi-variate vectors with uniform spectral measures $\Gamma(\cdot)$ and (\textit{b}) symmetric discrete spectral measures (located at intersections of axes with the unit sphere $S_2$).
 In (\textit{a}) noise induced jumps along $X$ and $Y$ axes are dependent. While in (\textit{b}) they are independent, however distribution of noise pulses is not elliptically symmetric.
 Situation (\textit{b}) is the only situation when components of the $\alpha$-stable vector are independent, i.e. their probability density can be factorized.
 Nevertheless, independence of noise induced jumps is not sufficient to assure independence of the variables $x$ and $y$. Usually, components of the position are dependent because deterministic force mixes variables. The only two exceptions are the motion of a free particle and a motion in a harmonic potential.

\section{Results \label{sec:results}}

First, equations for stationary states for 2D systems driven by bi-variate $\alpha$-stable noises are derived.
Next, using these equations some properties of stationary states are examined, see Sec.~\ref{sec:equations}.
Finally, main results are obtained by numerical simulations, see Sec.~\ref{sec:numericalsimulations}.

\subsection{Equations\label{sec:equations}}

Equations for 2D systems subject to the symmetric bi-variate $\alpha$-stable noise with uniform spectral measures and symmetric discrete spectral measures are derived. The continuous uniform spectral measure is constant on the unit sphere $S_2$, while
the symmetric discrete spectral measure is located at the intersections of the unit sphere $S_2$ with axes.

\subsubsection{Uniform spectral measures}

\begin{figure}[!ht]
\includegraphics[angle=0,width=0.98\columnwidth]{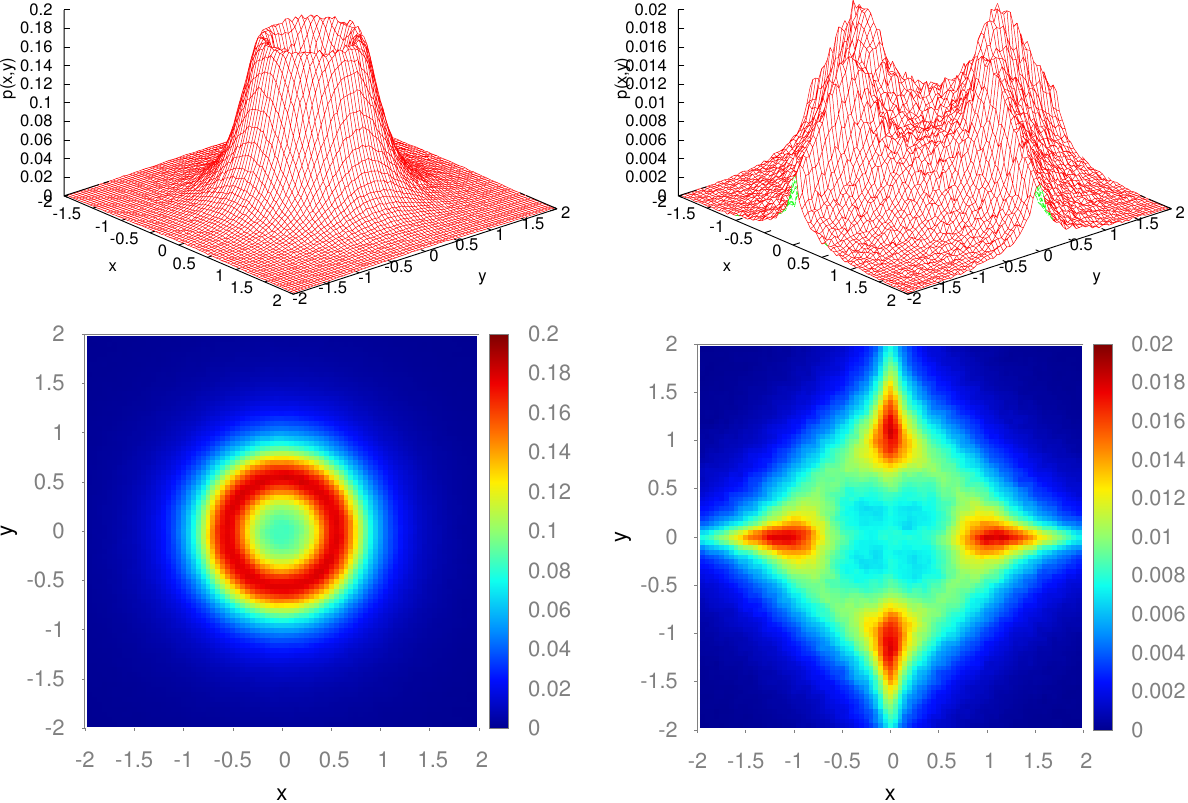}
 \caption{(Color online) Stationary states for the quartic potential $V(x,y)=\frac{1}{4}(x^2+y^2)^2$  subject to spherical (uniform)  $\alpha$-stable noise (left panel) and discrete $\alpha$-stable noise (right panel) with $\alpha=0.5$. Top row presents 3D surfaces while bottom row heat maps.
 The stationary densities have been estimated from the sample of $N=10^8$ elements with the integration time step $\Delta t=10^{-3}$.
}
\label{fig:quartic-a05}
\end{figure}

For the harmonic potential $V(x,y)=\frac{1}{2}(x^2+y^2)$ and uniform spectral measure the fractional diffusion equation is
\begin{equation}
 \frac{\partial p}{\partial t}= \frac{\partial }{\partial x} \left( x p \right) + \frac{\partial }{\partial y} \left( y p  \right) -( -\Delta)^{\alpha/2} p,
 \label{eq:harmonic}
\end{equation}
where $p=p(x,y,t)$. Additionally, for simplicity it has been assumed that $\sigma=1$.
In the Fourier space ($\mathcal{F}\left[f(\boldsymbol{r})\right]=\int e^{i\langle \boldsymbol{k},\boldsymbol{r} \rangle} f(\boldsymbol{r})d\boldsymbol{r}=\hat{f}(\boldsymbol{r})$ and $\boldsymbol{k}=[k,l]$) Eq.~(\ref{eq:harmonic}) attains the form
\begin{equation}
 \frac{\partial \hat{p}}{\partial t}=-k\frac{\partial \hat{p}}{\partial k} - l\frac{\partial \hat{p}}{\partial l} - (k^2+l^2)^{\nicefrac{\alpha}{2}} \hat{p}.
\end{equation}
The stationary density fulfills 
\begin{equation}
 k\frac{\partial \hat{p}}{\partial k} + l\frac{\partial \hat{p}}{\partial l} = -(k^2+l^2)^{\nicefrac{\alpha}{2}} \hat{p}.
 \label{eq:harmonic-fspace}
\end{equation}
Due to spherical symmetry, $p(x,y)=p\left(\sqrt{x^2+y^2}\right)$ and $\hat{p}(k,l)=\hat{p}\left(\sqrt{k^2+l^2}\right)$, Eq.~(\ref{eq:harmonic-fspace}) reduces to
\begin{equation}
 (k^2+l^2)^{\nicefrac{\alpha}{2}} \hat{p}+(k^2+l^2)^{\nicefrac{1}{2}} \hat{p}'=0,
 \label{eq:harmonic-reduced}
\end{equation}
where $\hat{p}'=\frac{\partial \hat{p}(\sqrt{k^2+l^2})}{\partial \sqrt{k^2+l^2}}$.
Eq.~(\ref{eq:harmonic-reduced}) has the solution
\begin{equation}
 \hat{p}=\exp\left[ - \frac{(k^2+l^2)^{\nicefrac{\alpha}{2}}}{\alpha} \right].
 \label{eq:harmonic-solution}
\end{equation}
According to \cite[Propositions~2.5.2 and~2.5.5]{samorodnitsky1994}, Eq.~(\ref{eq:harmonic-solution}) is the characteristic function of the bi-variate $\alpha$-stable density with the uniform spectral measure.
Therefore, in 2D like in 1D, the stationary state of the harmonic oscillator is the bi-variate $\alpha$-stable density like the one of the underlying noise. In particular, for $\alpha=2$, the stationary density is the bi-variate Gaussian distribution, while for $\alpha=1$ it is the bi-variate Cauchy distribution, see Eq.~(\ref{eq:spehricalcauchy}).

For the quartic oscillator driven by a $\alpha$-stable noise with uniform spectral measure situation is more complex.
The fractional diffusion equation is ($p=p\left(\sqrt{x^2+y^2}\right)$ and $\sigma=1$)
\begin{equation}
 \frac{\partial p}{\partial t}= \frac{\partial }{\partial x} \left[ (x^2+y^2)x p \right] + \frac{\partial }{\partial y} \left[ (x^2+y^2)y p  \right] -(- \Delta)^{\alpha/2} p.
 \label{eq:quartic}
\end{equation}
In the Fourier space 
\begin{equation}
 \frac{\partial \hat{p}}{\partial t}=k\frac{\partial^3 \hat{p}}{\partial k^3}
 + k\frac{\partial^3 \hat{p}}{\partial k \partial l^2}
+l\frac{\partial^3 \hat{p}}{\partial k^2 \partial l}
 + l\frac{\partial^3 \hat{p}}{\partial l^3} - (k^2+l^2)^{\nicefrac{\alpha}{2}} \hat{p}.
\end{equation}
The stationary density fulfills
\begin{equation}
 k\frac{\partial^3 \hat{p}}{\partial k^3}
 + k\frac{\partial^3 \hat{p}}{\partial k \partial l^2}
+l\frac{\partial^3 \hat{p}}{\partial k^2 \partial l}
 + l\frac{\partial^3 \hat{p}}{\partial l^3} = (k^2+l^2)^{\nicefrac{\alpha}{2}} \hat{p}.
 \label{eq:quartic-uniform-stat}
\end{equation}
Setting $z=\sqrt{k^2+l^2}$ one gets $\hat{p}(k,l)=\hat{p}(\sqrt{k^2+l^2})=\hat{p}(z)$ and
\begin{equation}
-\frac{\hat{p}'}{z}+\left[ \hat{p}''-z^{\alpha}\hat{p} \right]+z\hat{p}'''=0.
\label{eq:quartic-stationary}
\end{equation}
Eq.~(\ref{eq:quartic-stationary}) is associated with the following boundary conditions:
$\hat{p}(k,l)\big|_{(k,l)=(0,0)}=1$ (normalization),
$\partial_k\hat{p}(k,l)\big|_{(k,l)=(0,0)}=0$ (symmetry),
$\partial_l\hat{p}(k,l)\big|_{(k,l)=(0,0)}=0$ (symmetry),
$\partial^2_{kl}\hat{p}(k,l)\big|_{(k,l)=(0,0)}=0$  (symmetry) and $\lim_{k^2+l^2 \to +\infty}\hat{p}(k,l)=0$ (integrability).
For $\alpha=2$, the stationary state is $p(x,y)=\exp\left[ -\frac{(x^2+y^2)^2}{4}  \right]$, whose characteristic function obeys Eq.~(\ref{eq:quartic-stationary}).

\subsubsection{Discrete spectral measures}

For the harmonic potential $V(x,y)=\frac{1}{2}(x^2+y^2)$ and symmetric discrete spectral measure (concentrated on intersections of axes with the unit sphere $S_2$) the fractional diffusion equation is
\begin{equation}
 \frac{\partial p}{\partial t}= \frac{\partial }{\partial x} \left( x p \right) + \frac{\partial }{\partial y} \left( y p  \right) +
\left( \frac{\partial^\alpha}{\partial |x|^\alpha} + \frac{\partial^\alpha}{\partial |y|^\alpha}\right)p.
 \label{eq:harmonic-discrete}
\end{equation}
The stationary state of Eq.~(\ref{eq:harmonic-discrete}) can be found by the separation of  variables.
When $p(x,y)$ factorizes into $p(x)p(y)$ the Fourier transform factorizes into $\hat{p}=\hat{p}(k)\hat{p}(l)$.
The characteristic function (Fourier transform) satisfies
\begin{equation}
\hat{p}(l) \left[ k \frac{\partial \hat{p}(k)}{\partial k} - |k|^\alpha \hat{p}(k) \right]
+
\hat{p}(k) \left[ l \frac{\partial \hat{p}(l)}{\partial l} - |l|^\alpha \hat{p}(l) \right]
=0.
\end{equation}
Equations in square brackets are the same and their solution is just an 1D $\alpha$-stable density.
Consequently, due to symmetry of characteristic function the stationary state is a product of 1D $\alpha$-stable densities with the same parameters (except the scale parameter).
In particular, for $\alpha=1$ one gets product of two 1D Cauchy densities, see Eq.~(\ref{eq:discretecauchy}).

For the quartic potential the characteristic function of the stationary state fulfills
\begin{equation}
k\frac{\partial^3 \hat{p}}{\partial k^3}
 + k\frac{\partial^3 \hat{p}}{\partial k \partial l^2}
+l\frac{\partial^3 \hat{p}}{\partial k^2 \partial l}
 + l\frac{\partial^3 \hat{p}}{\partial l^3} - \left(|k|^\alpha+|l|^\alpha\right) \hat{p}=0
 \label{eq:quartic-discrete-stat}
\end{equation}
and equation does not separate like for the harmonic potential.
This is the natural consequence of the fact that despite independence of both noises the motion along both axes is no longer independent.

\subsection{Numerical Simulations\label{sec:numericalsimulations}}

Equations for the stationary densities, due to fractional derivatives, can be solved exactly only in a very limited number of cases.
Otherwise, one has to rely on numerical simulations.
Sec.~\ref{sec:tests} checks the correctness of implemented methods, while Sec.~\ref{sec:quartic} presents results of main simulations.

Stationary states for systems described by the fractional diffusion equation, see Eq.~(\ref{eq:ffpe2d}), can be estimated using Monte Carlo methods.
From large sample of trajectories generated according to the appropriate Langevin equation it is possible to estimate time dependent probability densities $p(x,y,t)$.
Stationary states are reached asymptomatically as $t\to\infty$, therefore it is necessary to perform sufficiently long simulations in order to assure that $p(x,y,t)$ density does not change any more.
Numerical simulations were performed with the integration time step $\Delta t=10^{-3}$ and averaged over $10^8$ repetitions.
Initially a test particle is located at the origin, i.e. $p(x,y,t)|_{t=0}=\delta(x)\delta(y)$. Analogously like in 1D, such an initial condition produces a slowly decaying cusp at the origin see \cite{chechkin2003,chechkin2004}. In forthcoming sections only stationary states are demonstrated.

\subsubsection{Tests\label{sec:tests}}

In order to verify correctness of implemented methods stationary states for parabolic and quartic potentials with bi-variate Gaussian ($\alpha=2$) noise and harmonic potential with bi-variate Cauchy ($\alpha=1$)  noises were constructed, see Figs.~\ref{fig:parabolic-quartic-gauss} and~\ref{fig:harmonic--cauchy}. For such a choice of parameters stationary states are known analytically, therefore results of simulations can be easily compared with exact formulas.

For multi-variate, uniform Gaussian noise and any bounding potential $V(\boldsymbol{r})$ the stationary density is of the Boltzmann-Gibbs type
\begin{equation}
 p_{\mathrm{st}}(\boldsymbol{r}) \propto \exp\left[ - \frac{V(\boldsymbol{r})}{\sigma^2}  \right],
\end{equation}
where $\sigma$ is the distribution width which depends on the scale parameter in the underlying Langevin and Smoluchowski-Fokker-Planck equations. Fig.~\ref{fig:parabolic-quartic-gauss} presents results for the bi-variate uniform Gaussian white noise with parabolic $V(x,y)=\frac{1}{2}(x^2+y^2)$ (left panel) and quartic $V(x,y)=\frac{1}{4}(x^2+y^2)^2$ (right panel) potentials. Results are presented both as 3D surfaces (top row) and heat maps (bottom rows). The constructed stationary densities perfectly agree with appropriate Boltzmann-Gibbs densities.

\begin{figure}[!ht]
\includegraphics[angle=0,width=0.98\columnwidth]{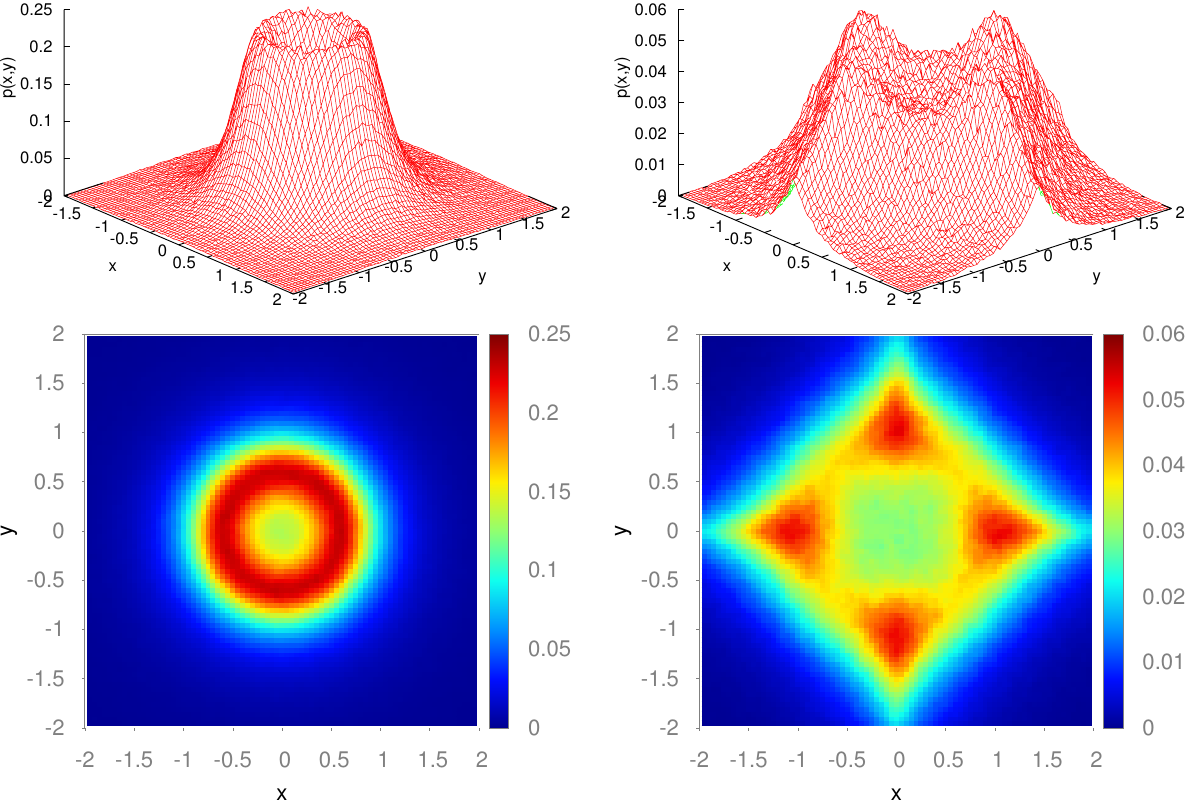}
 \caption{(Color online) Stationary states for the quartic potential $V(x,y)=\frac{1}{4}(x^2+y^2)^2$  subject to spherical (uniform)  $\alpha$-stable noise (left panel) and discrete $\alpha$-stable noise (right panel) with $\alpha=0.7$. Top row presents 3D surfaces while bottom row heat maps.
 The stationary densities have been estimated from the sample of $N=10^8$ elements with the integration time step $\Delta t=10^{-3}$.
}
\label{fig:quartic-a07}
\end{figure}

Figure~\ref{fig:harmonic--cauchy} shows stationary states for the bi-variate Cauchy oscillator subject to spherically symmetric (uniform) (left panel) and discrete (right panel) bi-variate Cauchy noise. For the spherical symmetric, uniform Cauchy noise the stationary state is bi-variate spherically symmetric Cauchy distribution, see Eq.~(\ref{eq:spehricalcauchy}) and left panel of Fig.~\ref{fig:harmonic--cauchy}. This is in accordance with the general prediction, that for the harmonic potential and bi-variate $\alpha$-stable noise with uniform spectral measure, the stationary density is of the same shape as the one of the noise. Finally, the right panel of Fig.~\ref{fig:harmonic--cauchy} demonstrates stationary states for symmetric, discrete bi-variate Cauchy noise, i.e. the $\alpha$-stable noise with a symmetric, discrete spectral measure located on the intersections of axes with the unit sphere $S_2$. In such a case, displacements along both axes are independent. Consequently, the stationary state is the
product of two
Cauchy distributions, see Eq.~(\ref{eq:discretecauchy}), and again the stationary state agrees with the underlying noise distribution (up to the scale parameter).

\begin{figure}[!ht]
\includegraphics[angle=0,width=0.98\columnwidth]{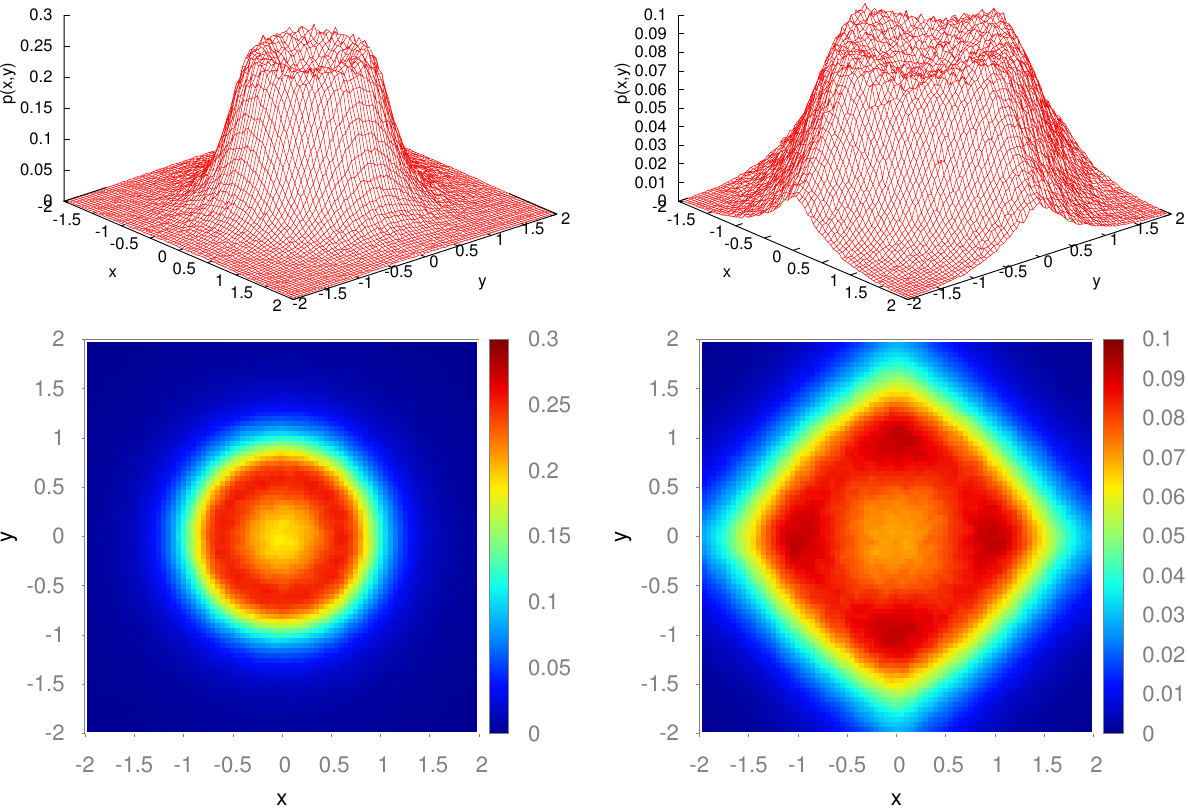}
 \caption{(Color online) Stationary states for the quartic potential $V(x,y)=\frac{1}{4}(x^2+y^2)^2$  subject to spherical (uniform)  $\alpha$-stable noise (left panel) and discrete $\alpha$-stable noise (right panel) with $\alpha=1.0$, i.e. the Cauchy noise. Top row presents 3D surfaces while bottom row heat maps.
 The stationary densities have been estimated from the sample of $N=10^8$ elements with the integration time step $\Delta t=10^{-3}$.
}
\label{fig:quartic-a10}
\end{figure}

\begin{figure}[!ht]
\includegraphics[angle=0,width=0.98\columnwidth]{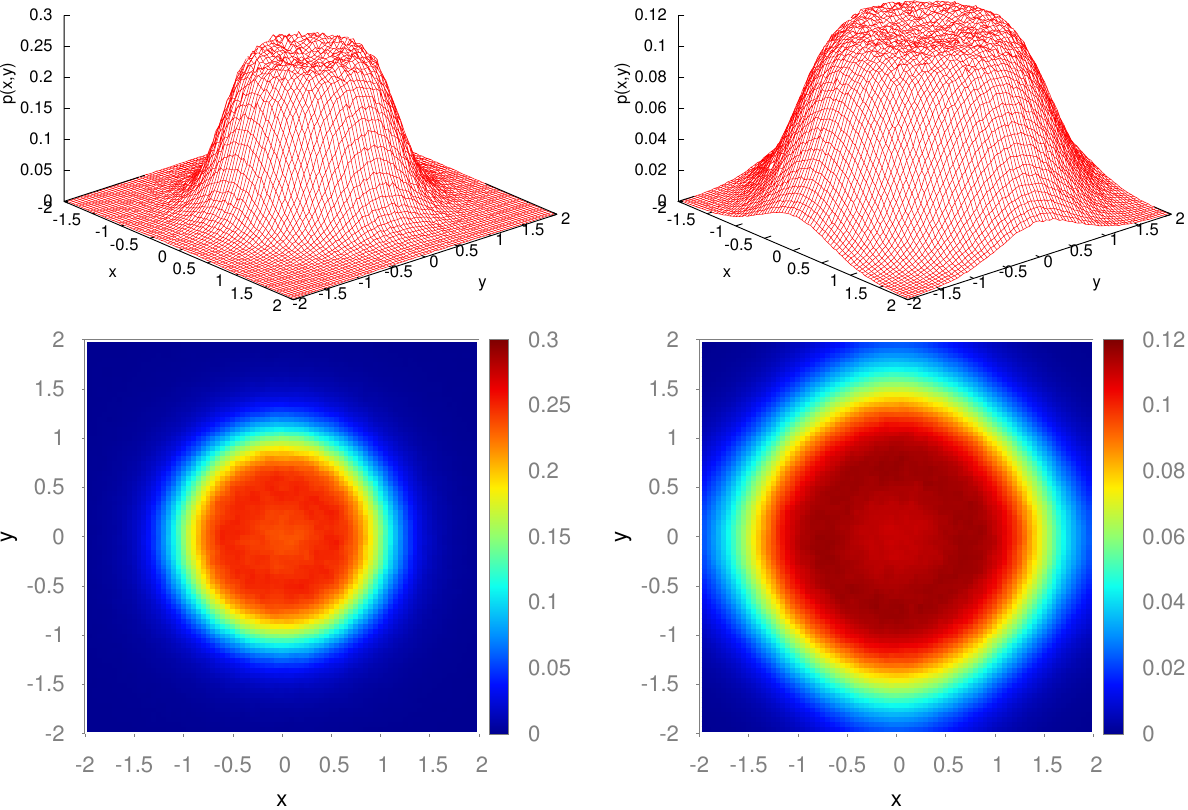}
 \caption{(Color online) Stationary states for the quartic potential $V(x,y)=\frac{1}{4}(x^2+y^2)^2$  subject to spherical (uniform)  $\alpha$-stable noise (left panel) and discrete $\alpha$-stable noise (right panel) with $\alpha=1.5$. Top row presents 3D surfaces while bottom row heat maps.
 The stationary densities have been estimated from the sample of $N=10^8$ elements with the integration time step $\Delta t=10^{-3}$.
}
\label{fig:quartic-a15}
\end{figure}

\subsubsection{Quartic potential\label{sec:quartic}}

In the 1D case, the stationary state is also known for the quartic Cauchy oscillator, see Eq.~(\ref{eq:1d-quartic}).
Nevertheless, the bi-variate, symmetric, discrete quartic Cauchy oscillator cannot be used for testing purposes because jumps along both axes are no longer independent. For the quartic potential, the deterministic force $\boldsymbol{F}(x,y)=-\nabla V(x,y)=-[(x^2+y^2)x,(x^2+y^2)y]$ mixes coordinates and consequently the diffusion equation cannot be solved by the separation of variables. Therefore, stationary states cannot be expressed as a product of stationary densities for a corresponding 1D potential.

Following figures~\ref{fig:quartic-a05}--\ref{fig:quartic-a19} show results of simulations of the quartic $\alpha$-stable oscillator for a limited choice of bi-variate $\alpha$-stable noises: with symmetric continuous, uniform spectral measures (left columns) and symmetric, discrete spectral measures concentrated on intersections of axes with the unit sphere $S_2$ (right panels). The presented stationary densities have pronounced minimas at $(0,0)$.

Stationary states for the spherically symmetric $\alpha$-stable noise are spherically symmetric. More precisely, symmetries of the stationary density reflect symmetries of spectral measures. In particular for uniform spectral measures, stationary states are spherically symmetric, i.e. they depend on $r=\sqrt{x^2+y^2}$, see left panels of Figs.~\ref{fig:quartic-a05}--\ref{fig:quartic-a19}. For $\alpha<2$, stationary densities have well defined minima at the origin which are surrounded by maxima located at the circle. For the discrete symmetric spectral measures, see right panels of Figs.~\ref{fig:quartic-a05}--\ref{fig:quartic-a19}, stationary states are no longer spherically symmetric. Nevertheless, stationary densities still have pronounced minima at the origin and maxima on axes. With the increasing $\alpha$, both for uniform continuous and symmetric discrete spectral measures, minima become shallower.  Moreover as $\alpha$ approaches 2, both stationary densities become similar. Finally, in the limiting 
case of $\alpha=2$, 
they are exactly the same.

\begin{figure}[!ht]
\includegraphics[angle=0,width=0.98\columnwidth]{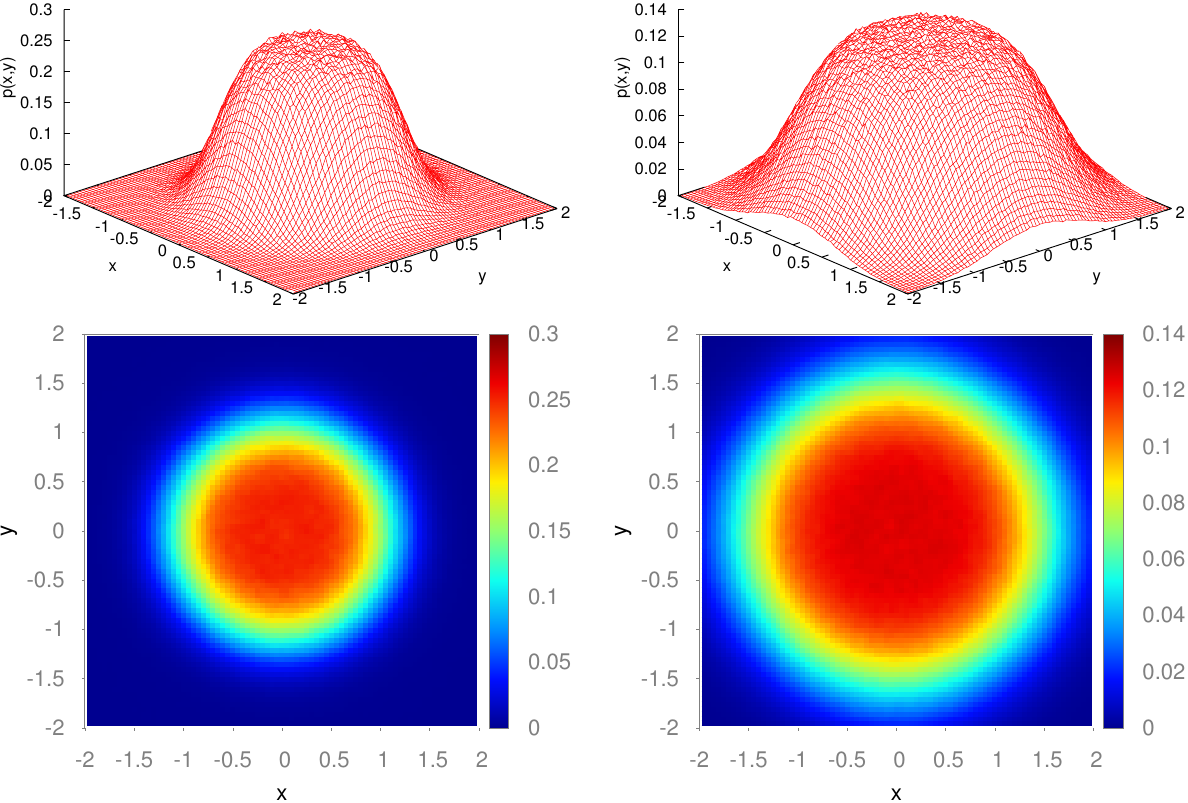}
 \caption{(Color online) Stationary states for the quartic potential $V(x,y)=\frac{1}{4}(x^2+y^2)^2$  subject to spherical (uniform)  $\alpha$-stable noise (left panel) and discrete $\alpha$-stable noise (right panel) with $\alpha=1.9$. Top row presents 3D surfaces while bottom row heat maps.
 The stationary densities have been estimated from the sample of $N=10^8$ elements with the integration time step $\Delta t=10^{-3}$.
} 
\label{fig:quartic-a19}
\end{figure}

\begin{figure}[!ht]
\includegraphics[angle=0,width=0.98\columnwidth]{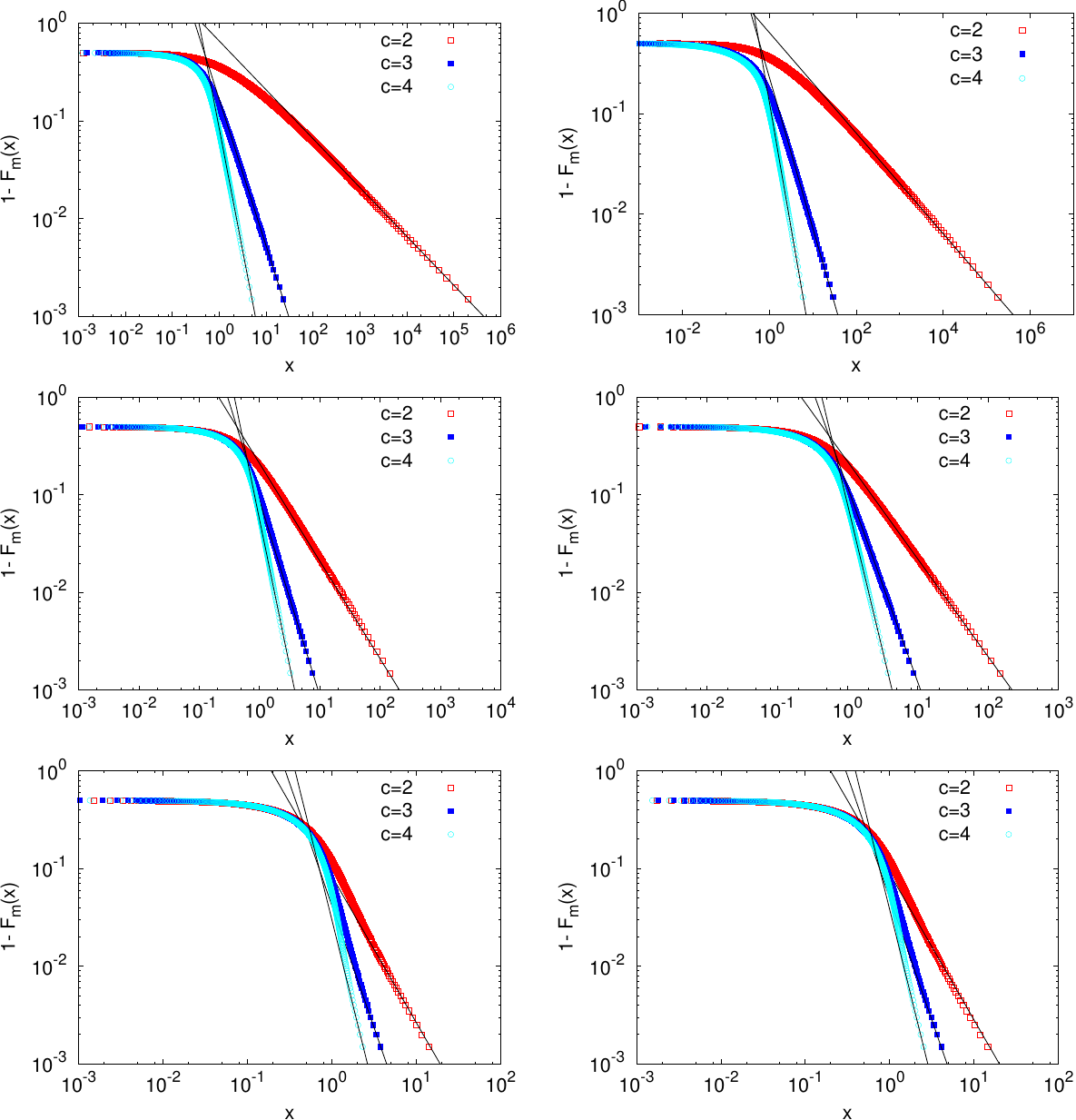}
 \caption{(Color online) Survival probabilities, $S(x)=1-F_m(x)$, for marginal densities of $x$ for systems  perturbed by bi-variate $\alpha$-stable noises  with continuous uniform (left panel) and symmetric discrete (right panel) spectral measures. Various rows correspond to various values of the stability index $\alpha$: $\alpha=0.5$ (top row), $\alpha=1$ (middle row) and $\alpha=1.5$ (bottom row). Different curves correspond to various potentials of $V(x,y)=(x^2+y^2)^{c/2}$ type:  harmonic ($c=2$), cubic ($c=3$) and quartic ($c=4$).
 Solid lines present  $x^{-(c+\alpha-2)}$ power-law asymptotics of survival probailities.
 The stationary densities have been estimated from the sample of $N=10^7$ elements with the integration time step $\Delta t=10^{-3}$.
}
\label{fig:quartic-cs}
\end{figure}

Finally, Fig.~\ref{fig:quartic-cs} examines asymptotics of marginal densities of $x$ for stationary states, i.e. $p(x)=\int p(x,y)dy$, for the general single well potentials of $(x^2+y^2)^{{c}/{2}}$ type with $c=2$ (harmonic),  $c=3$ (cubic) and $c=4$ (quartic) potentials  subject to bi-variate $\alpha$-stable noises with continuous uniform (left panel) and symmetric discrete (right panel) spectral measures. Various rows correspond to various values of the stability index $\alpha$: $\alpha=0.5$ (top row), $\alpha=1$ (middle row) and $\alpha=1.5$ (bottom row).
For uniform spectral measures, due to spherical symmetry of stationary states, all cross sections are the same and they have $(x^2+y^2)^{-{(c+\alpha)}/{2}}$ asymptotics. For discrete spectral measures cross sections along axes are the same and they display $|x|^{-(c+\alpha-1)}$ or  $|y|^{-(c+\alpha-1)}$ asymptotics.
The marginal probability densities for systems driven by bi-variate $\alpha$-stable noises with continuous uniform and discrete symmetric spectral measures have the same asymptotics, i.e.  $p(x) \propto  |x|^{-(c+\alpha-1)}$ and $p(y) \propto  |y|^{-(c+\alpha-1)}$. This is corroborated by Eqs.~(\ref{eq:quartic-uniform-stat}) and~(\ref{eq:quartic-discrete-stat}) which lead to the same equations for marginal densities in both cases  resulting in the same asymptotics.
The asymptotics of marginal densities is better visible when instead of marginal densities survival probabilities, i.e. $S(x)=1-F_m(x)=1-\int\limits_{-\infty}^x p(x)dx$, are depicted, see Fig.~\ref{fig:quartic-cs} which confirms hypothesis regarding asymptotics of stationary states.
The observed asymptotic behavior could be anticipated from the fractional Smoluchowski-Fokker-Planck equation, see the discussion below Eq.~(\ref{eq:ffpe2d}), in the analogous way like for the 1D systems \cite{chechkin2003,chechkin2004,chechkin2006}.

\section{Summary and Conclusions\label{sec:summary}}

1D systems driven by $\alpha$-stable noises display some unexpected properties, which are very different from characteristics of systems driven by white Gaussian noise.
First of all, stationary states exist for potential wells which are steep enough.
Otherwise, the distribution width grows in time.
Secondly, in the harmonic potential the stationary state reproduce the probability density associated with the underlying noise (except the scale parameter).
Finally, for potential wells stepper than parabolic stationary states can be multimodal.

Main properties of 1D stationary states of systems driven by $\alpha$-stable noise transfer into 2D realms.
However, special attention is required because bi-variate $\alpha$-stable densities are determined by the spectral measure.
Various spectral measures result in very different properties of bi-variate $\alpha$-stable noises.
Consequently, the 2D systems driven by bi-variate $\alpha$-stable noises are described by the whole family of Langevin equations and associated fractional Smoluchowski-Fokker-Planck equations depending on the choice of the spectral measure.

For the 2D harmonic potential and continuous uniform or discrete, symmetric located on intersections of axes with the unit sphere spectral measures stationary states reconstruct (up to rescaling) the noise distributions. In the limit of $\alpha=2$ bi-variate $\alpha$-stable densities converge to bi-variate Gaussian distribution. Therefore, both types of bi-variate $\alpha$-stable noises produces the same stationary states.

For the quartic potential stationary states have local minima at the origin both for uniform and symmetric, discrete spectral measures. Additionally, for uniform spectral measures stationary states are spherically symmetric. With increasing value of the stability index $\alpha$ minima become shallower. Finally, for $\alpha=2$ the Boltzmann-Gibbs distribution is reconstructed.

In general for single well potentials of $(x^2+y^2)^{c/2}$ (with $c \geqslant 2$) subject to action of bi-variate $\alpha$-stable noises with uniform spectral measures stationary densities have power-law  $(x^2+y^2)^{-{(c+\alpha)}/{2}}$ asymptotics. Consequently, marginal densities have also power-law asymptotics with the exponent ${-(c+\alpha-1)}$. For the discrete symmetric spectral measures concentrated on the intersection of the unit sphere with axes stationary states are no longer spherically symmetric. Nevertheless, marginal densities are of the same asymptotics like for uniform spectral measures.
Finally, one can expect that stationary states exist also for subharmonic potentials with large enough exponent $c$.
Using normalization condition one can speculate that, analogously like in 1D, exponent characterizing steepness of the potential should be $c>2-\alpha$.
Nevertheless, this issue requires further verification.


%


\begin{acknowledgments}
Computer simulations have been performed at the Academic
Computer Center Cyfronet, Akademia G\'orniczo-Hutnicza (Krak\'ow, Poland) under CPU grant
MNiSW/Zeus\_lokalnie/UJ/052/2012.

\end{acknowledgments}


\end{document}